\documentclass[thmsa,12pt,onecolumn,notitlepage]{article}

\begin{document}
\begin{flushright}
BARC-TIFR/CM/98/301(I)\\
To be presented at "Experimental Workshop on High Temperature Superconductors and Related Materials 
(Advanced Activities)" to be held from October 19$^{th}$ to November 6$^{th}$, 1998 in S. C. Bariloche, 
Argentina.\\
\end{flushright}

\begin{center}
\small
\smallskip {\large Evidence of a first order like jump in equilibrium
magnetization across the peak effect region in superconducting 2H-NbSe$_2$}
\end{center}

\smallskip

\begin{center}
G. Ravikumar$^{*}$, P. K. Mishra and V. C. Sahni

Technical Physics and Prototype Engineering Division, Bhabha Atomic Research
Centre, Mumbai - 400085, India

S. S. Banerjee, S. Ramakrishnan and A. K. Grover$^{*}$

Tata Institute of Fundamental Research, Homi Bhabha Road, Mumbai - 400005,
India

\smallskip

P. L. Gammel, D. J. Bishop and E. Bucher

Lucent Technologies, Murray Hill, NJ 07940, USA

S. Bhattacharya

NEC Research Institute, Princeton, NJ 08540, USA

Abstract
\end{center}

{\scriptsize We report magnetization hysteresis measurements in the peak
effect region of a very clean single crystal of superconducting 2H-NbSe$_2$ at
temperatures very close to T$_c$(0). Through measurement of minor magnetization
curves, we infer the equilibrium magnetization curve across the PE region.
We observe a small first order like change in equilibrium magnetization across the peak effect. We relate this 
observation to the entropy change
associated with the order-disorder transformation. We further note that the first order like change in the
equilibrium magnetization observed in 2H-NbSe$_2$ is comparable to the change(s) observed
in the equilibrium magnetization at the FLL\ melting transition in various
high temperature superconductors.}

\smallskip

The advent of high T$_c$ superconductivity focused wide spread attention on
pristine physics issue of melting of pure (pinning free) Abrikosov flux line
lattice (FLL)[1]. There is now a growing and compelling evidence from magnetic[2-4],
thermal[5] and structural studies[6,7] in a variety of 
superconductors, that the vortex solid to vortex liquid transformation is of
first order (or weakly first order)[2-4]. The crucial input in favor of this has
stemmed from the observation of a step increase in the equilibrium
magnetization $(\Delta M_{eq})$. At temperature values where FLL melting is
anticipated, the Clausius- Clayperon equation,

$L=T$ $\Delta S=-T$ $\Delta M_{eq}$($dH_m/dT)$~~~~~~~~~~(1)

\noindent provides the necessary connection between the latent heat $L$ and $%
\Delta M_{eq}$ via the knowledge of the slope of the FLL melting transition (%
$H_m,T$) curve. However, the step increase in the equilibrium magnetization
has been experimentally discerned only across limited ( and different) $H-T$
regions in crystals of BiSrCaCuO [2] and YBaCuO [3]. In
LaSrCuO system it could however be observed over a much wider $(H,T)$ region[4].
Keeping in view that FLL melting is a generic phenomenon related to collapse
of long range spatial order of FLL, its fingerprint(s) ought to be
observable even in the mixed state of conventional low temperature
superconductors across an appropriate ($H,T$) region. However, there are
fewer reports of such claims in conventional low T$_c$ compounds as compared
to those in high T$_c$ cuprates. FLL melting in conventional superconductors
is difficult to investigate experimentally, because it is believed to occur
very close to $H_{c2}$[1], where there could be additional complications
arising from the fluctuations in the phase of the superconducting order
parameter as well.

Two prerequisites[1] for facilitating the observation of FLL melting in a
specimen of conventional superconductor are: (i) a large value of the
Ginzburg number G$_i=(1/2)(kT/H_c^2\xi ^3\epsilon )^{2}$ ; and (ii) the existence of
appreciable pinning free region (where magnetization is reversible) below
superconductor-normal phase boundary. Anisotropic hexagonal 2H-NbSe$_2$
system with $T_c(0)$ = 7.2K has a $G_i$ $\sim$ 10$^{-4}$[8]; this value lies in between those
of high T$_c$ cuprates (10$^{-2}$) and low T$_c$ alloy
superconductors (10$^{-6}$). This system is therefore considered appropriate
to explore the phenomenon of FLL\ melting in the domain of conventional low T%
$_c$ alloy superconductors. As stated above FLL\ melting is a pure system concept
and in cuprate superconductors its fingerprint in reversible region of dc
magnetization data is convenient to locate only in clean single crystal
samples which are free of structural defects. The high purity single
crystals of 2H-NbSe$_2$ system rank favorably amongst the most weakly
pinned superconducting samples of all varieties of superconducting systems.
The ratio $J_c$/$J_0$ (where $J_c$ is the critical current density and $J_0$
is the depairing current density) in clean crystals of 2H-NbSe$_2$ is
typically $\sim $10$^{-6}$[8] and this value is orders of magnitude lower than
those usually observed in conventional low T$_c$ superconductors. However
all high quality single crystals of 2H-NbSe$_2$ display the peak effect (PE)
phenomenon[9], which is the anomalous increase in $J_c$ close to $H_{c2}(T)$
where FLL\ melting is expected. Ever since the initial impetus injected by
an explanation of PE phenomenon by Pippard[10] in terms of rapid collapse of
rigidity of vortex lattice, various (other ) possible close connections
between FLL melting and PE\ phenomenon have been provided from time to time
via static and dynamical measurements[8] on vortex states as well as via
theoretical treatments and simulation studies[11,12].

In isothermal dc magnetization measurements, the peak effect manifests as an anomalous increase in
the magnetization hysteresis. It is understood that the width of the
hysteresis loop at a given field H provides a measure of $J_c(H)$. Recent
structural studies through small angle neutron experiments in single
crystals of Nb[6] and $\mu $SR\ measurements in 2H-NbSe$_2$ [7] support the
existence of an intimate relationship between the reduction in spatial order
of FLL\ and the PE\ phenomenon. The entropy change associated with the collapse
of long range order in the FLL therefore needs to be discerned from the change(s) in
the equilibrium magnetization across the PE\ region. We report in this paper
a successful outcome of our attempts to locate a step change in the
equilibrium magnetization ($\Delta M_{eq}$) across the peak effect region
from the dc magnetization hysteresis data in a clean crystal of 2H-NbSe$_2$.

Isothermal magnetization data (across PE region) has been measured using a Quantum
Design SQUID magnetometer with field parallel to $c$-axis of a clean 2H-NbSe$
_2$ single crystal, at $T$ = 6.85K and 6.95K respectively. The data at
T=6.95K is recorded with a 2 cm scan. However at 6.85K, the magnetization
hysteresis is comparable to the field inhomogeneity along the scan length thereby necessitating
the use of half-scan technique [13] with an effective 4 cm scan. This particular crystal is very weakly
pinned and the PE can be observed down to a field
of 50 Oe in temperature dependent ac susceptibility ($\chi$$^{\prime}$) experiment (see inset of Fig.1. for 
$\chi$$^{\prime}$(T) plot in H = 1kOe, however all data not
shown here). The locus of peak temperature $T_p$ vs $H$ in it has the
behavior (Fig.1) which lies in between those for single crystals A and B of
2H-NbSe$_2$ studied by us earlier (see Fig.4 of Ref [9]). In isothermal dc magnetization hysteresis data (see 
Figs 2(a) and
2(b)), the PE region has been identified with the anomalous increase in the
hysteresis just below H$_{c2}$ values. The peak fields $H_p$ at 6.85K and
6.95K are consistent with $T_p(H)$ curve shown in Fig.1. It may be noted
that the forward and reverse legs of the hysteresis envelope in the PE
region are significantly asymmetric. Further, the field value $H_{pl}^+$ at
which anomalous increase in the diamagnetic response commences on the
forward magnetization curve differs significantly from the field $%
H_{pl}^{-}(<H_{pl}^+)$ where PE\ ends on the reverse leg.

In clean single crystals of high T$_c$ superconductors, the dc magnetization
is reversible over a wide range of fields prior to H$_{c2}$ and any
modulation (step change, inflection points etc.) in equilibrium
magnetization can therefore be identified distinctly. However, when the
magnetization is irreversible, $M_{eq}$ values are usually obtained[14] as,

$M_{eq}(H)=[M^{+}(H)+M^{-}(H)]/2$,~~~~~~~~~~~(2)

\noindent where $M^{+}\,$and $M^{-}$ are the magnetization values measured
in ascending and descending field cycles. Each of these values comprises
contributions from shielding currents set up in the sample in addition to
the equilibrium magnetization. An implicit assumption in this relation is
that the critical current density $J_c$ at a given $H$ remains the same on ascending
and descending fields. In other words $J_c$ is independent of magnetic history
of the vortex state. In recent years transport[15], dc magnetization and
ac susceptibility[16] studies have revealed that $J_c$ in weakly pinned
superconducting samples (which show PE phenomena) could strongly depend on
their thermomagnetic history. While studying the effect of thermomagnetic
history on transport critical currents in a crystal of Niobium which showed
PE, Steingart et al [17] had noted the inequality,

$J_c^{FC}(H)>J_c(H^{-})>J_c(H^{+})$,~~~~~~~~~(3)

\noindent for fields below $H_p$ (see Fig.2a for identification of peak field H$_p$). $%
J_c^{FC}(H)$ is the critical current density in field cooled (FC) state and $%
J_c(H^{-})$ and $J_c(H^{+})$ are the critical current densities measured in
decreasing and increasing fields respectively. 
Originally, Steingart et al[17] had surmised that the vortex state in the field
cooled state is most strongly pinned as each of the vortex lines attempts to
conform to maximum number of pinning sites as the flux lines nucleate below $%
H_{c2}$. In view of the above inequality, Eq.(2)
cannot be used for obtaining $M_{eq}$ for $H < H_p$ and a suitable alternative has to be found so that the 
pair of magnetization values in Eq.2 correspond to the same value of $J_c(H)$. However, for $H>H_p$, 
where $J_c$ is
observed to be independent of magnetic history, $M_{eq}$ could be obtained using Eq.2.

Within Larkin-Ovchinikov collective pinning description[18] (Larkin volume $V_c$ = $R_c^2 L_c$ 
$\propto$ $J_c^{-2}$, where $R_c$ and $L_c$ are radial and longitudinal correlation lengths respectively) 
the inequality $%
J_c(H^{-})>J_c(H^{+})$ implies that the extent of order in the vortex state
generated in decreasing field (across PE region) from above $H_{c2}$ is less
than that in the vortex state created while increasing the field from zero
value. In other words, the so called Larkin volume at a given field, over which FLL remains
correlated, is larger on the increasing field cycle as compared to that on
the decreasing field cycle. In an isothermal dc magnetization experiment,
this inequality (which holds upto $H_p$) results in a hysteresis loop
which is asymmetric (as shown in Fig.2a and 2(b)) with respect to the equilibrium magnetization because
the contribution of the induced shielding current to the magnetization on
the reverse leg would be larger in magnitude than that on the forward leg.
Considering that the PE phenomena relates to reduction in spatial order of
the vortex array, the observation $H_{pl}^+ >H_{pl}^{-}$ implies that the fully
disordered state, occurring at $H=H_p$ on the decreasing field cycle, does not fully heal back to the ordered
state as field values are reduced. In fact the healing process continues at least down to $%
H=H_{pl}^{-}$. We believe that the existence of path dependence, i.e., $H_{pl}^{-}$ $%
\neq $ $H_{pl}^+\,$,is another manifestation of first order nature of the
order-disorder transformation accompanying the PE phenomenon.

Roy and Chaddah[19] have proposed that the minor magnetization curves obtained
by reversing the field from different values lying on the forward
magnetization curve could be used to construct the requisite reverse leg of
the hysteresis curve which is a symmetric counterpart of the forward
magnetization curve[14]. Within Critical State model, such minor hysteresis
curves merge into the reverse magnetization curve. However, in the PE\
region of CeRu$_2$, minor magnetization curves initiated from a field $H$
lying between $H_{pl}$ and $H_p,$ saturate without merging into the usual
descending envelope curve. They [19] have used the saturated value $%
M_{minor}^{-}(H)$ in Eq.1 instead of $M^{-}(H)\,$to obtain $M_{eq}(H)$. We
also show in Figs.2(a) and 2(b), the magnetization curves obtained while reversing the
field from different values lying on the forward magnetization curve at
6.85K and 6.95K in our crystal of 2H-NbSe$_2$. We note that the minor magnetization curves originating
from field values lying between H$_{pl}$ and H$_p$ do not reach upto the usual reverse
magnetization curve, whereas those originating from $H>H_p$ merge with
the reverse envelope curve which is consistent with the observations of in CeRu$_2$ [19] . In addition, we 
observe that for certain range of fields below $%
H_{pl}$ also, the minor curves saturate without merging with reverse magnetization
curve. We would like to assert that these minor curves are the symmetric counterparts (i.e., notionally 
correspond to same $J_c(H)$) of the forward magnetization curve and correspond to the more ordered FLL 
compared to that on the reverse curve. Following Ref [19], we use the saturated magnetization M$^-
_{minor}$ on the minor magnetization curve in Eq.1 to obtain $M_{eq}$. The more ordered FLL can 
alternatively be generated starting from a field cooled vortex lattice (for H $<$ H$_{pl}$) and subjecting it 
to an increasing (FC-FOR) and decreasing (FC-REV) fields. Since the current density in FC state is much
larger than those in the increasing and decreasing fields,
the minor hysteresis loops initially overshoot [16] the envelope hysteresis
curves as shown in Fig.3. FC-FOR curve eventually merges into the forward magnetization curve,  whereas 
the FC-REV curve merges into the minor magnetization curve initiated from the forward magnetization 
curve.
It is pertinent
to remark here that the minor magnetization curve (at 6.85 K) initiated from the forward curve merges into 
the usual reverse (obtained by reversing from H $>$ H$_{c2}$) magnetization curve at fields
sufficiently below $H_{pl}$. Also, it may be pointed out
that if the field is increased from the (reverse) minor hysteresis curve, the magnetization readily merges into 
the usual forward envelope
curve.
 
We understand the observed behavior of the minor magnetization
curves within the LO collective pinning description [18] by the following
argument. For a small decrement in field from H lying on the forward
magnetization curve, the shielding currents merely reverse sign while the
magnitude is maintained same as that on the forward magnetization curve. The size of
Larkin domains or the extent of FLL\ correlations essentially remain unaltered. For
the same field value H, on the reverse magnetization curve, the sign of the
induced currents are same as those on the minor magnetization curve but are
of larger magnitude. As stated earlier, a locus of the $M_{minor}^{-}(H)$ values obtained at
different fields (see dotted curve in Fig.2a) appears reasonably symmetric
with respect to the usual forward magnetization curve. Thus, if we use $
M_{minor}^{-}$ instead of $M^{-}$ in Eqn.1, we obtain the $M_{eq}$ as shown in
the inset of Fig.2. The step increase in $M_{eq}\,$across the PE region can therefore be
easily identified (see $\Delta M_{eq}$ as marked in the inset of Fig 2(a)). The tiny peak
like modulation in $M_{eq}$ values between $H_{pl}\,$and $H_p$ is reminiscent
of similar behavior across the PE region in some samples of CeRu$_2$ as
reported in Ref. 19. It has been argued that the change in $M_{eq}$ at $H_{pl}$ in the case of CeRu$_2$ is 
an imprint of a first order onset of formation of a new superconducting phase with enhanced pinning. 
However, we attribute this $\Delta M_{eq}$ to
the onset of amorphisation of FLL as a consequence of thermal softening of
its elastic modulii across the PE\ region. Latent heat across the PE region
in 2H-NbSe$_2$ can be estimated by substituting the value of $
dH_{pl}/dT$ ($\approx$ $dH_p/dT$)$\approx$-5$\times 10^{3}$ Oe/K (Fig.1) along with $\Delta 
M_{eq}$ = 380mOe and 230 mOe at 6.95K and 6.85K, respectively. Our values of $\Delta M_{eq}$ and 
$L$ compare favorably with similar
estimates across FLL melting transition in cuprate superconductors[2-4] and
across the PE region in CeRu$_2$.

To conclude, we have presented an estimate of step change in equilibrium
magnetization ($\Delta M_{eq}$) extracted from dc magnetization hysteresis
across the PE region at 6.85 K and 6.95K at $H_p$ 1.7 kOe and 1.0 kOe, respectively in a single crystal of
2H-NbSe$_2$ for $H_{dc}$ parallel to $\,c$. From a variety of
detailed transport measurements in pure crystals of 2H-NbSe$_2$ which
elucidated the dynamics of driven vortex matter prior to and across the PE
region, it had been argued that \cite{r8} the vortex matter at peak position of PE
region is in a {\it pinned liquid} state. Two recent structural studies via $\mu $
SR experiments [17] in this system have established that the spatial order of FLL
undergoes a sudden change at the onset of PE in temperature dependent scans
in 500 $<$ $H$ $<$ 200 Oe. Though PE region extends
over 200 Oe in Fig.2(a) and 2(b), the peak effect phenomenon implies a sharp
transformation in the state of vortex matter. The transition width of PE in
temperature dependent ac susceptibility measurements (at fixed $H$) is
smaller than the width of the normal to superconducting transition in zero
field [9]. Thus, we believe that our estimate of  $\Delta $ $M_{eq}$ across PE
reliably determines the latent heat associated with the occurrence of order
to disorder transformation across PE. We find that the estimated value of $%
\Delta M_{eq}$ in 2H-NbSe$_2$ is of the same
order as those observed earlier at FLL melting transition in crystals of
cuprate superconductors [2-4]. We also note that our values of $\Delta M_{eq}$ in
2H-NbSe$_2$ also compare favorably with those reported across the PE\ region in some samples of
superconducting CeRu$_2$[19].

We would like to acknowledge fruitful discussions with Nandini Trivedi, Satya Majumder and Mahesh 
Chandran.
\vskip 0.2truecm
$^{*}$ Corresponding authors.\\
e-mails: gurazada@apsara.barc.ernet.in or grover@tifr.res.in
\begin{enumerate}
\item  G. Blatter, M. V. Feigel'man, V. B. Geshkenbein, A. I. Larkin and V. M. Vinokur, Rev.Mod.Phys. 66 
(1994) 1125 and references therein.

\item  E. Zeldov et al, Nature (London) 375 (1995) 373

\item  U. Welp et al, Phys.Rev.Lett. 76 (1996) 4809 

\item  T. Sasagawa et al, Phys.Rev.Lett. 80 (1998)4297 and references therein.

\item  A. Schilling et al, Nature 382 (1996) 791

\item  P. L. Gammel et al, Phys.Rev.Lett. 80 (1998) 833.

\item  T. V. Chandrasekhar Rao et al, Physica C 299 (1998) 267; Phys. Rev. Lett. (submitted) 

\item  S. Bhattacharya and M. J. Higgins, Physica C 257 (1996) 232 and references therein.

\item  S. S. Banerjee et al, Physica C (in press)

\item  A. B. Pippard, Philos. Mag. 19 (1969) 217.

\item  A. I. Larkin, M. C. Marchetti, V. M. Vinokur, Phys.Rev.Lett 75 (1995) 2992

\item  C. Tang, X. Ling, S. Bhattacharya, P. M. Chaikin, Europhys.Lett 35(1996) 597 and references therein.

\item  G. Ravikumar et al Physica C 276 (1997) 9; 298 (1998) 122.

\item  P. Chaddah, S. B. Roy , Mahesh Chandran, Phys. Rev.B (in press)

\item  W. Henderson, et al Phys.Rev.Lett. 77 (1996) 2077

\item  G. Ravikumar et al, Phys.Rev.B 57 (1998) R11069

\item  M. Steingart, A. G. Putz, E. J. Kramer, J.Appl.Phys.44 (1973) 5580

\item  A. I. Larkin and Y. N. Ovchinikov, J. Low Temp Phys. 34 (1979) 409.

\item  S. B. Roy and P. Chaddah, J.Phys: Condens. Matter 9 (1997) L625; 10
(1998) 4885; Physica C 273 (1996) 120; Phys.Rev.B 55 (1997) 11100.
\end{enumerate}

{\normalsize \textbf{\textsl{\pagebreak }}}

Figure 1: A comparison of Peak Effect curve (locus of T$_p$ vs H) in 2H-NbSe$_2$ crystals used in the 
present study with those in the crystal A and B studied in Ref[9]. The dotted lines identify PE curves in 
crystals in A and B, whereas the filled circle data points in the present sample corresponds to peak 
temperatures T$_p$(H) as determined from ac susceptibility measurements shown in the inset. The T$_p$ 
values in different crystals have been normalized to the respective T$_c$(0) values. For the sake of 
completeness of the magnetic phase diagram the H$_{c1}$(T) and H$_{c2}$(T) curves have also been 
included in the main panel. \\
\vskip 0.5 truecm
\noindent
Figure 2(a): Isothermal magnetization hysteresis data for H$\parallel$c across the Peak Effect region in 2H-
NbSe$_2$ at 6.85 K. H$_{pl}$$^+$ and H$_{pl}$$^-$ identify the field values at which PE notionally 
commences and terminates along the usual forward and reverse hysteresis paths. H$_p$ identifies the peak 
field of PE. It also includes the minor hysteresis curves generated by reversing the fields from H = 1600 Oe 
and 1640 Oe and 1740 Oe lying on the forward hysteresis leg. The minor curves initiated from H = 1600 Oe 
($<$ H$_{pl}$) and 1640 Oe (H$_{pl}$ $<$ H $<$ H$_p$) do not reach upto the reverse envelope curve, 
instead they lie just above the forward envelope curve. The dotted curve passing through the saturated 
values of M$_{minor}$$^-$ (see text or Ref.19) sketches the new reverse envelope curve which appears 
more symmetric with respect to the forward envelope curve as compared to the usually experimentally 
measured reverse envelope curve by decreasing the field from above H$_{c2}$. The equilibrium 
magnetization values prior to PE (i.e., for H $<$ H$_{pl}$) lie in between the forward envelope curve and 
new reverse envelop curve, whereas the equilibrium value above the PE (H $>$ H$_{irr}$) can be readily 
identified with the (path independent) measured (magnetization value. The inset shows the step change in 
equilibrium magnetization ($\Delta$M$_{eq}$) across the PE region. Two straight line have been drawn in 
the inset to guide the eye about the occurrence of $\Delta$M$_{eq}$ across the PE region.\\
\vskip 0.5 truecm
\noindent
Figure 2(b): Isothermal magnetization hysteresis data for H$_{dc}$ $\parallel$ c across PE region in 2H-
NbSe$_2$ at 6.95 K. The identity of different symbols in this figure is the same as described in the caption 
of Fig.2(a). The inset shows the step change in equilibrium magnetization $\Delta$M$_{eq}$ across the PE 
region.
\vskip 0.5 truecm
\noindent
Figure 3: Minor magnetization curves generated by decreasing (REV) or increasing (FOR) the field from 
field cooled magnetization value at H = 1300 ($<$ H$_{pl}$) at 6.85 K. The FC - REV and FC - FOR 
curves initially overshoot the respective envelope curves thereby showing that J$_c$$^{FC}$ is much larger 
than the J$_c$ values along the envelope curves. It is to be noted that FC-FOR curve readily merges into the 
forward envelope curve whereas the FC-REV curve drops down to values which lie significantly below the 
usually measured reverse envelope. In fact FC-REV curve appears to lie very close to the forward envelope 
curve, thereby implying very low values of J$_c$ (for H $<$ H$_{pl}$) for vortex states on the forward 
envelope curve.

\end{document}